
\overfullrule=0pt
\font\eightit=cmti8
\outer\def\beginsection#1\par{\vskip0pt plus.3\vsize\penalty-100
   \vskip0pt plus-.3\vsize\bigskip\vskip\parskip
   \message{#1}\leftline{\bf#1}\nobreak\smallskip}
\def\da{{\cal D}(A)}
\def\db{{\cal D}(B)}
\def\iq{\pi_{\cal Q}^*}
\def\IR{{\rm I\kern -1.6pt{\rm R}}}
\def\IP{{\rm I\kern -1.6pt{\rm P}}}
\def\ZZ{{\rm Z\kern -4.0pt{\rm Z}}}
\def\IC{\ {\rm I\kern -6.0pt{\rm C}}}
\def\O{{\Omega}}
\def\gs{\left(\matrix{0\cr1\cr}\right)}
\def\pq{\pi_{\cal Q}^{\phantom.}}
\def\Rt{R_\theta}
\def\rt{{R_{\theta(t)}}}
\def\ta{{\bf \tau}}
\let\hat=\widehat
\def\sqr#1#2{{\vcenter{\vbox{\hrule height.#2pt
\hbox{\vrule width.#2pt height #1pt \kern#1pt
\vrule width.#2pt}
\hrule height.#2pt}}}}
\def\square{\mathchoice\sqr56\sqr56\sqr{2.1}3\sqr{1.5}3}

\def\pmb#1{\setbox0=\hbox{$#1$}%
\kern-.025em\copy0\kern-\wd0
\kern.05em\copy0\kern-\wd0
\kern-.025em\raise.0433em\box0 }

\def\pmbb#1{\setbox0=\hbox{$\scriptstyle#1$}%
\kern-.025em\copy0\kern-\wd0
\kern.05em\copy0\kern-\wd0
\kern-.025em\raise.0433em\box0 }

\hbox to \hsize{\hfil\eightit EACEHL-Revised July-12-92}
\vglue1.5truein
\centerline{\bf OPTIMAL HYPERCONTRACTIVITY FOR FERMI FIELDS AND}
\centerline{\bf RELATED NON-COMMUTATIVE INTEGRATION
INEQUALITIES}
\bigskip

{\baselineskip = 12pt
\halign{\qquad#\hfil\qquad\qquad\hfil&#\hfil\cr
Eric  A. Carlen\footnote{$^*$}{On leave from School of Math., Georgia
Institute of Technology, Atlanta, GA 30332} & Elliott H.
Lieb\footnote{$^{**}$}{Work supported by U.S.
National Science Foundation grant no. PHY90--19433--A01.}\cr
Department of Mathematics & Departments of Mathematics and Physics\cr
Princeton University & Princeton University\cr
Princeton, New Jersey  08544 & Princeton, New Jersey  08544--0708\cr}}
\bigskip

\vskip .4 true in
\centerline{Dedicated to Prof. Huzihiro Araki on his 60$^{th}$ birthday}
\vskip 1 true in
{\bf Abstract:}  Optimal
hypercontractivity bounds for the fermion oscillator semigroup are obtained.
These are the fermion analogs of the optimal hypercontractivity bounds for
the boson oscillator semigroup obtained by Nelson.
In the process, several results of independent interest in the theory of
non-commutative integration are established.
\vfill\eject

\def\J{{\cal J}}
\def\Q{{\cal Q}}
\def\P{{\cal P}}
\def\K{\cal K}
\def\cq{{\cal C}({\cal Q})}
\def\cqn{{\cal C}({\cal Q}_{(n-1)})}
\def\cqp{{\cal C}^p({\cal Q})}
\def\cqq{{\cal C}^q({\cal Q})}
\def\cqt{{\cal C}^2({\cal Q})}

\def\ck{{\cal C}({\cal K})}
\def\ckp{{\cal C}^p({\cal K})}
\def\ckt{{\cal C}^2({\cal K})}
\def\H{{\cal H}}

\centerline{\bf I. INTRODUCTION}
\vskip .3 true in

Observables pertaining to the configuration of a quantum system
with $n$ degrees of freedom are
operators $Q_1, Q_2,\dots, Q_n$ which, depending on the system,
may or may not commute.
Our main concern is with the case in which the configuration variables are
amplitudes of certain field modes.

For boson fields, these configuration observables
do commute,
and the state space $\H$ can
be taken as the space of all complex square integrable
functions on their joint spectrum. This is
the Schr\"odinger $q$-space representation, and the fact that in it the
state space is a function space, and not just an abstract Hilbert space, is
very helpful in the analysis of such systems. As one example, it
sometimes turns out that physically interesting operators preserve the cone of
positive functions, and this opens the way to the application of the
Perron-Frobenius theorem in the study of ground states of
such systems.

For fermion fields, the configuration observables do not commute,
and this simple $q$-space representation is not available. However, the
non-commutative integration theory of Irving Segal [Se53]
permits the construction
of a suitable substitute, and in fact it was created with such a purpose in
view. This approach to the study of fermion systems has been extensively
developed by Gross [Gr72] who, among other things, proved a version of the
Perron-Frobenius theorem adapted to the setting and applied it to
prove existence and uniqueness of ground states for certain fermion
quantum field models.

The main estimate which enabled Gross to apply his Perron-Frobenius type
theorem to fermion fields was a hypercontractivity estimate for the fermion
oscillator semigroup.
Corresponding hypercontractivity estimates for boson fields had been
introduced earlier by Nelson [Ne66] who later obtained the optimal
such bound for bosons [Ne73].

Our main result, Theorem 4 below,
is the corresponding optimal hypercontractivity bound for
fermions. Before stating this theorem, we describe its mathematical and
physical contexts in some detail because it cannot even be formulated
naturally in the conventional Fock space language.
(Of course its perturbation theoretic
{\it consequences} can be expressed quite naturally in the usual
language.)

The state space for a system of $n$ fermion degrees of freedom is
conventionally realized as the
Fock space
$${\cal F} =
\oplus_{j=o}^n\bigl((\IC^n)^{\wedge j}\bigr)\eqno(1.1)$$
The basic ``free Hamiltonian'' on this space is the fermion number
operator
$$\hat H_0 = \sum_{j=1}^n c^*_jc_j\eqno(1.2)$$
where $c_j$ and $c_j^*$ are the usual fermion annihilation and creation
operators acting on ${\cal F}$. The fermion oscillator semigroup is
the semigroup of operators $\exp(-t\hat H_0)$ that it generates.

For $n$ boson degrees of freedom, the state space may be realized
{\it either} as
the boson Fock space, or as $L^2(\IR^n,(2\pi)^{n/2}e^{-x^2/2}{\rm d}x)$. The
natural isomorphism between these spaces was pointed out by
Segal, and the latter may be regarded as the Schr\"odinger $q$-space
realization. The boson oscillator semigroup is the semigroup generated by the
boson number operator. Though we have just defined it in Fock space terms,
it may also be considered as an operator semigroup on
$L^2(\IR^n,(2\pi)^{n/2}e^{-x^2/2}{\rm d}x)$. In this setting, it
is {\bf hypercontractive}; i.e., for any finite $p$
greater than 2, there is $t_p$ suffciently
large, for which the semigroup is a contraction from
$L^2(\IR^n,(2\pi)^{n/2}e^{-x^2/2}{\rm d}x)$
to $L^p(\IR^n,(2\pi)^{n/2}e^{-x^2/2}{\rm d}x)$ for all $t\ge t_p$. Since
$t_p$ is independent of $n$, this result proved useful in treating
perturbation theoretic problems for boson fields. This hypercontractivity
inequality cannot be formulated naturally in the Fock space setting because
no natural notion of ``$L^p$'' can be introduced there. The
$L^p(\IR^n,(2\pi)^{n/2}e^{-x^2/2}{\rm d}x)$ setting is essential.

In order to to pass from the Fock space description
for a system of $n$ fermion degrees of freedom to a non-commutative
analog of the
Schr\"odinger $q$-space description (in which we have non-commutative
analogs of $L^p$ norms),
we introduce configuration observables
$Q_j =  c_j + c_j^*$, and let $\cq$ be the algebra with unit which
they generate. This is an operator algebra on the $2^n$--dimensional Hilbert
space ${\cal F}$. As we explain in Section II, it is a Clifford algebra
naturally associated to $\IC^n$ with its usual inner product.

What follows here shall be explained in more detail in Sections II--IV,
but the key fact for our present considerations is that the $2^n$
distinct monomials in the $Q_j$ are a basis for $\cq$ as a
{\it vector space}.
Thus, any $A$ in $\cq$ can be uniquely written as
$$A = \alpha I + \sum_{j}\alpha_jQ_j + \sum_{j<k}\alpha_{j,k}Q_jQ_k +
\cdots + \alpha_{1,\dots,n}Q_1\cdots Q_n\quad.\eqno(1.3)$$
It is not hard to see that any vector $v$ in ${\cal F}$ can be written
as $v = A\Omega$ for some unique $A\in\cq$ acting on the vacuum $\Omega$.
The operator $\exp(-t\hat H_0)$ acts on such vectors $v$. Thus,
$\exp(-t\hat H_0)v = B\Omega$ for some
unique $B\in\cq$ (which depends on $t$ and $A$,
of course). The map $P_t:A\mapsto B$ is clearly linear, and a simple
calculation (elucidated in Section IV) shows that
$$P_tA = \alpha I + e^{-t}\sum_{j}\alpha_jQ_j +
e^{-2t}\sum_{j<k}\alpha_{j,k}Q_jQ_k +
\cdots + e^{-nt}\alpha_{1,\dots,n}Q_1\cdots Q_n\quad.\eqno(1.4)$$

Instead of considering Fock space, ${\cal F}$, and the operator
$\exp(-t\hat H_0)$, Segal's idea is to consider $\cq$ and the linear operator
$P_t$ that acts on $\cq$ as above.
Following Segal, we can construct $L^p$ norms on $\cq$, the most
important case being $p=2$, in terms of which one can do quantum
mechanics without mentioning Fock space.

To construct the norms,
define the linear functional $\tau:\cq\rightarrow \IC$ by
$$\tau(A) = \langle\Omega, A\Omega\rangle_{\cal F}\quad.\eqno(1.5)$$
It turns out, as we explain in Section III that $\tau(A)$ is,
for $A\in \cq$, nothing other
than the usual normalized trace on ${\cal F}$ applied to $A$.

The basic ingredients of a non-commutative integration theory are an algebra
of operators such as $\cq$, and a linear functional on it such as $\ta$.
Norms analogous to the $L^p$ norms of commutative integration theory can
now be introduced by
$$\|A\|_p = \bigl(\ta(A^*A)^{p/2})\bigr)^{1/p}\qquad
{\rm for}\qquad 1\le p <\infty\quad;$$
$\|A\|_\infty$ denotes the operator norm of $A$.
Let  $\cqp$
denote $\cq$ equipped with the corresponding norm.

Having expanded our horizons beyond Hilbert space, we can ask for bounds
between $\|P_tA\|_q$ and $\|A\|_p$ for different $q$, $p$ and $t$
Our main result, Theorem 4, is
the optimal {\bf fermion hypercontractivity} inequality; i.e.
for all $1<p\le q<\infty$ and all $A$ in $\cqp$,
$$\|P_tA\|_q \le \|A\|_p\qquad{\rm when}\qquad e^{-2t} \le {p-1\over
q-1}\quad,\eqno(1.6)$$
and the $t$ saturating the inequality on the right
is the smallest for which the inequality
on the left always holds. We prove this for $n$ degrees of freedom
with $n$ an
arbitrary finite integer. Since the estimate is independent of $n$, a
theorem of Gross [Gr72]
implies that it holds as well with infinitely many degrees
of freedom.

The result (1.6) was conjectured by Gross [Gr75]
who proved it [Gr72] in the special case
$p=2$, $q=4$. The cases $p=2$, $q=2m$
where $m$ is an integer were proved by Lindsay and Meyer [LiMe]
following earlier
work by Lindsay [Lin] on the case $p=2$, $q=2^m$. Since $P_t$ is self adjoint,
duality yields a corresponding family of results in which $q=2$. Until now
all other cases (except when $n=1\ {\rm or}\ 2$) had remained open.
The optimal relation between $t$, $p$ and $q$ found here for fermion
hypercontractivity is the same as that found by Nelson [Ne73] for boson
hypercontractivity.

By now there are many proofs of Nelson's inequality. Neveu's elegant proof
[Nev],
like Nelson's original proof, is based on probabilistic methods. Proofs
based on geometric methods have been given by Carlen and
Loss [CL90] and by Lieb [Li90] who considers generalizations in which the
Mehler kernel (the kernel for the boson oscillator semigroup) is replaced by an
arbitrary Gaussian kernel. More references can be found in the
bibliography to Gross's article [Gr89]. However, none of the existing
approaches to the boson problem has been found to solve the
fermion problem.

As a corollary to the optimal fermion hypercontractivity inequality,
we obtain the optimal
{\bf fermion logarithmic Sobolev inequality}:
$$\ta\bigl(|A|^2\ln|A|^2\bigr)
- \bigl(\|A\|_2^2\ln\|A\|_2^2\bigr) \le 2\ta(A^*H_0A)
\eqno(1.7)$$
where $|A| = (A^*A)^{1/2}$.

This inequality was conjectured by Gross, and proved by him
in a weaker form
with the constant on the right increased by a factor of $\ln3$.
In studying perturbations of $H_0$ by multiplication operators $V$, this
inequality plays the same role as does the usual Sobolev inequality in studying
perturbations of $-{1\over 2}\Delta$ by multiplication operators $V$ on
$L^2(\IR^n,{\rm d}^nx)$ [Fe69].
(The multiplication operator associated to a self-adjoint element
$V$ of $\cq$ is defined to be the average of left and right
multiplication by $V$, and is denoted here by the same symbol $V$.)
Again, in the standard Fock space setting, there is no natural way to
formulate such an {a-priori} regularity inequality, or even to introduce
the notion of a multiplication operator.

This paper is organized as follows: In Section II we study the
structure of $\cq$ for finite $n$. It is actually simpler, as well as
technically advantageous, to consider it as a subalgebra of the algebra
$\ck$ generated by the identity, the configuration observables
$Q_1,\dots,Q_n$ and their conjugate momenta $P_1\dots,P_n$.
Since they both turn out
to be certain Clifford algebras, their structure has been worked out
long ago with the representation theory of the orthogonal groups. Thus, this
section contains no new result but simply introduces notation and
prepares the way for what follows. Of particular use are an explicit spin--
chain representation of the operator algebra $\ck$, and the Jordan-Wigner
transform identifying it with the algebra generated by $n$ ``hard core
bosons''.

Section III concerns properties of the spaces $\ckp$ and their
norms. The main result here is an optimal uniform convexity inequality
for $\ckp$, $1 < p \le 2$, which is joint work with Keith Ball. We need
only a special case of this inequality here,
and a proof is provided in an appendix
for the reader's convenience.

Section IV introduces a convenient expression for $P_t$ in
terms of the conditional expectation $\pq$ of $\ck$ with respect to
$\cq$.
The main result in this section is an inequality for conditional
expectations which
enables us to prove that
$$\sup\{\|P_tA\|_q\ : \ \|A\|_p=1\}\qquad=\qquad
\sup\{\|P_tA\|_q\ : \ A\ge 0\ {\rm and}\ \|A\|_p=1\}\quad.$$
Thus, to establish (1.6) in general, we need only consider
positive $A$.

In Section V we establish the optimal fermion hypercontractivity bounds
and the corresponding optimal fermion logarithmic Sobolev inequality.
This is done in several steps. First, using results collected
in Sections II and III we establish that (1.7) holds for $1 < p \le 2$
and $q=2$. At $t=0$ this is an equality; differentiating it there yields
the logarithmic Sobolev inequality (1.7).
Gross showed that (1.6) would follow from (1.7), if it were true (as we
show here),
for all {\it self adjoint} $A$.
His result rests on a deep inequality he established for positive operators.
Since it is {\it not} in general true that
$\|P_tA\|_q \le \|P_t|A|\|_q$, as is trivially true in the commutatative
case, Gross's result does not allow us to conclude (1.6) for {\it general}
(i.e. non-self adjoint) $A$ from (1.7).
The results of the Section IV do, however, allow us to draw this
conclusion.

Finally, in Section VI we show that the same hypercontractivity relation
(1.6) holds for a mixed system of bosons and fermions.

Non-commutative probability theory has grown into a substantial branch of
analysis with a number of physical applications. The mathematical theory is
reviewed and developed in [Me85] and [Me86],
while other sorts of physical applications,
besides those discussed here, are treated in [Da76] and [HuPa] for
example.

It is a pleasure to thank Leonard Gross
for discussing his results and conjectures
with us, and for encouraging us to take up the latter. We are indebted to
Keith Ball for his collaboration on the subject of convexity inequalities
that led to Theorem 1 [BCL], which is one of the key ingredients in the
present work. Thanks are also
due to G.-F. Dell'Antonio, A. Jaffe and A. Wightman for useful discussions.

\vskip.3 true in
\centerline{\bf II. FERMIONS AND THE CLIFFORD ALGEBRA}
\vskip.3 true in

We begin by recalling for later use some well known facts about fermions.

The fundamental observables for a system of $n$ fermion
degrees of freedom are {\bf configuration
operators} $Q_1, Q_2,\dots , Q_n$ together with their {\bf conjugate momenta
operators}
$P_1, P_2,\dots, P_n$ all acting as operators on a complex Hilbert space
$\H$ and satisfying the {\bf canonical anticommutation relations}:
$$Q_jQ_k + Q_kQ_j = 2\delta_{jk}\quad,\quad P_jP_k + P_kP_j =
2\delta_{jk}\quad,\quad
P_jQ_k + Q_kP_j = 0\quad.\eqno(2.1)$$

We denote the complex algebra generated by the identity and
the configuration observables
by $\cq$,
and the complex algebra generated by the identity, the configuration
observables
and the momentum observables all together by $\ck$.
$\cq$ is the object of primary interest; but many
aspects of its structure are most readily seen within the larger algebra
$\ck$.

This algebra can be concretely represented as the algebra of observables for
a spin--chain as follows. We define the matrices
$$ I = \left[\matrix{1&0\cr 0&1\cr}\right]\quad,\quad
U = \left[\matrix{1&0\cr 0&-1\cr}\right]\quad,\quad
Q = \left[\matrix{0&1\cr 1&0\cr}\right]\quad,\quad
P = \left[\matrix{0&i\cr -i&0\cr}\right]\quad.$$

Let $\H$ denote the $n$-fold tensor product of $\IC^2$ with itself:
$$\H = \underbrace{\IC^2\otimes\dots\otimes\IC^2}_{n\rm\;times}\quad,$$
and on $\H$ define the operators
$$Q_j = U\otimes\cdots U\otimes Q\otimes I\otimes\cdots\otimes I\quad,\quad
P_j = U\otimes\cdots U\otimes P\otimes I\otimes\cdots\otimes I\eqno(2.2)$$
where the $Q$ and the $P$ occur in the $j$th places.

The operators $Q_1,\dots,Q_n,P_1\dots,P_n$ just defined are easily seen to
satisfy the canonical anticommutaion relations. Of great use in studying
the algebra $\ck$ that they generate is the fact that it is also the algebra
generated by $n$ ``hard core boson'' degrees of freedom. More explicitly,
put
$$\hat Q_j = I\otimes\cdots I\otimes Q\otimes I\otimes\cdots\otimes I\quad
,\quad
\hat P_j = I\otimes\cdots I\otimes P\otimes I\otimes\cdots\otimes I\quad,
\eqno(2.3)$$
and call the algebra generated by the operators
$\hat Q_1,\dots,\hat Q_n,\hat P_1\dots,\hat P_n$ the {\bf hard core boson
algebra}, and denote it by
$\hat{\cal C}({\cal K})$.
To see that the two algebras coincide, put
$$U_j = I\otimes\cdots I\otimes U\otimes I\otimes\cdots\otimes I\qquad
{\rm and}\qquad
V_k = \prod_{j=1}^{k-1}U_j\quad.\eqno(2.4)$$
Then since
$P_jQ_j = \hat P_j\hat Q_j = iU_j$,
each $V_k$ belongs to both the hard core boson algebra and $\ck$.
Moreover
$$\hat Q_k = V_k Q_k\quad{\rm and}\quad\hat P_k = V_kP_k\eqno(2.5)$$
with the inverse relation given as well by
$$Q_k = V_k\hat Q_k\quad{\rm and}\quad P_k = V_k\hat P_k\quad.\eqno(2.6)$$
Thus, $Q_k$,$P_k$ are in
$\hat{\cal C}({\cal K})$, and
$\hat Q_k$,$\hat P_k$ are in $\ck$.

What we call the hard core boson algebra was initially introduced by
Jordan and Klein [JoKl] as a first attempt to implement the
Pauli exclusion principle mathematically.
The transformation of observables (2.6)
was discovered by
Jordan and Wigner [JoWi] and used by them to write the algebraic relations
characterizing the algebra in the familiar covariant form (2.1); today
it is known as the Jordan--Wigner transform.
It has been used many times since --
for example, in a solution of the two dimensional
Ising model by Schultz, Mattis and Lieb, whose paper [SML] can be consulted
for references to other applications. It is also
the key to Brauer and Weyl's treatment [BrWe] of the Clifford algebra
on which the exposition in the rest of this section is largely based.
Following them,
we now show that
$\ck$ is the full matrix algebra on $\H$.

First, we introduce a basis in $\H$.
Let
$$ e_1 = \left(\matrix{1\cr 0\cr}\right)\qquad{\rm and}\qquad
e_{-1} = \left(\matrix{0\cr 1\cr}\right)$$
be the standard basis of $\IC^2$.
For each $j =
1,\dots, n$, let $\sigma_j$ be either $1$ or $-1$. Then the unit vectors
$$e_{\sigma_1,\dots,\sigma_n} = e_{\sigma_1}\otimes\cdots \otimes
e_{\sigma_n}$$
provide a natural orthonormal basis for $\H$.

Next introduce the fermion creation
and annihilation operators
$c^*_j = {1\over 2}(Q_j - iP_j)$ and $c_j = {1\over 2}(Q_j +
iP_j)$
and their hard core boson analogs
$\hat c^*_j = {1\over 2}( \hat Q_j - i\hat P_j)$ and
$\hat c_j = {1\over 2}(\hat Q_j +
i\hat P_j)$.
Now put
$L(\sigma_1,\dots,\sigma_n) = \prod_{j=1}^nB_j$
where
$B_j = \hat c_j$ if $\sigma_j = 1$ and
$B_j = \hat c_j\hat c_j^*$ if $\sigma_j = -1$.
Then
$L(\sigma_1,\dots,\sigma_n)
e_{\sigma_1,\dots,\sigma_n} = \O$ where
$$\O = \underbrace{\gs\otimes\cdots\otimes \gs}_{n\rm\; times}
= e_{-1,-1,\dots,-1}\eqno(2.7)$$
is a distinguished unit vector in $\H$ called the {\bf ground state}. Also,
$L(\sigma_1,\dots,\sigma_n)$ annihilates all the other basis vectors.
Moreover,
$L^*(\sigma_1,\dots,\sigma_n)\Omega =
e_{\sigma_1,\dots,\sigma_n}$.
Thus,
$$L^*(\ta_1,\dots,\ta_n)
L(\sigma_1,\dots,\sigma_n)
e_{\sigma_1,\dots,\sigma_n} =
e_{\ta_1,\dots,\ta_n}\quad,$$ and this operator
annihilates all other basis vectors.
Manifestly this operator belongs to
$\hat{\cal C}({\cal K})$, and
hence to $\ck$ as well, and the $2^n$ operators of this kind form a basis
for the full matrix algebra on ${\cal H}$.

This concrete description of $\ck$ in terms of spin--chain observables is the
most useful for many purposes. Still, it is also useful to have a
characterization of $\ck$ which is less dependent on co\"ordinates; i.e.
on the choice of fermi configuration observables $Q_1,\dots,Q_n$.

Toward this end, consider the standard $n$--dimensional Hilbert space
$\IC^n$ equipped with its standard inner product $(\cdot,\cdot)$
and complex conjugation.
Let  $\K$ denote $\IC^n$ considered as a real $2n$--dimensional Hilbert space
equipped with the inner product
$$\langle x,y\rangle_{\K}^{\phantom .} = \Re(x,y)\quad.$$
Then complex conjugation on $\IC^n$ induces an
involutory orthogonal transformation
$J$ on $\K$. Let $\Q$ and $\P$ respectively denote the eigenspaces
of $J$ corresponding to the eigenvalues
$+1$ and $-1$.
The bilinear form on $\K$ given by
$\Im(x,y)$
is symplectic so that $\K$ is naturally endowed with the structure
possessed by the classical phase space of a system of $n$ linear degrees of
freedom. $\Q$ is called the {\bf configuration space}, $\P$ the
{\bf momentum space},
and the complex conjugation $J$ is usually identified with time reversal.

The Clifford algebra $\ck$ is characterized up to automorphism as the
algebra with unit $I$ such that:

\item{(i)}\ There is a linear imbedding $\J:\K \rightarrow \ck$, and
$\J(\K)$ generates $\ck$.

\item{(ii)}\ For all $x,y \in \K$,
$$\J(x)\J(y) + \J(y)\J(x) = \langle x,y\rangle_{\K}^{\phantom .}
\quad.\eqno(2.8)$$

To make contact with our previous concrete description, let
$\{q_1,\dots,q_n\}$ be an orthonormal basis of\ $\IC^n$ consisting of purely
real vectors. For each $j$, let $p_j = iq_j$. Evidently ${\cal Q}$ is
spanned by
$\{q_1,\dots,q_n\}$ and ${\cal P}$ is spanned by
$\{p_1,\dots,p_n\}$. Any $x\in{\cal K}$ can then be written as
$x= \sum_{j=1}^n\xi_jq_j + \sum_{j=1}^n\eta_jp_j$.
Using the notation introduced above, put $\J:\K\rightarrow
\ck$ by
$$\J(x) =
\sum_{j=1}^n\xi_jQ_j + \sum_{j=1}^n\eta_jP_j\quad.$$

Let $\{x_1,\dots,x_{2n}\}$ be any orthonormal basis of $\ck$. Then the
monomials
$$\J(x_{\alpha_1})\J(X_{\alpha_2})\cdots\J(x_{\alpha_k})$$
together with $I$ form a basis for the algebra. It is easy to see that
the product of any two such monomials is a third. Though
the multiplication rule can be simply
expressed in terms of certain contraction rules, its precise form
is not useful to us here. What is useful to observe is
that since the right side of (2.8) is invariant under orthogonal
transformations of $\K$, the multiplication law of these monomials
does not depend on the choice of the orthonormal basis. For this reason,
any orthogonal transformation $R$ of $\K$ induces an automorphism of $\ck$
which we shall also denote by $R$.
Indeed, with
$R:\ck \rightarrow \ck$ defined by
$$R\bigl(\J(x_{\alpha_1})\cdots\J(x_{\alpha_k})\bigr) =
\J(R(x_{\alpha_1}))\cdots\J(R(x_{\alpha_k}))\quad,$$
$R$ is evidently invertible and
by the remarks made just above one easily sees that for all $A,B \in \ck$,
$R(AB) = R(A)R(B)$ which is to say that $R$ is an automorphism.

Finally, we remark that $\ck$ is a $*$-algebra; there is a unique
conjugate linear involutory antiautomorphism $A\mapsto A^*$ which is the
identity on $\J(\K)$. It is given by
$$R\bigl(\J(x_{\alpha_1})\cdots\J(x_{\alpha_k})\bigr)^* =
\J(x_{\alpha_k})\cdots\J(x_{\alpha_1})\quad.$$
Of course, on $\ck$ regarded as a matrix algebra, this is just the usual
adjoint. Evidently the automorphism $R$ of $\ck$ induced by an orthogonal
transformation $R$ of $\K$ is a $*$-automorphism; i.e. $R(A^*) =
\bigl(R(A)\bigr)^*$.

The facts that orthogonal transformations of
$\K$ induce automorphisms of $\ck$, that $\ck$ is a full matrix algebra, and
that all automorphisms of full matrix algebras are inner; i.e. of the form
$A\mapsto SAS^{-1}$ for some nonsingular matrix $S$, are the basis of
Brauer and Weyl's treatment of the spin representations of the orthogonal
groups [BrWe]. We will use the fact that all automorphisms of $\ck$ are
inner
several times in what follows.
\vskip .3 true in
\centerline{\bf III. ANALYSIS ON THE CLIFFORD ALGEBRA}
\vskip .3 true in

Let ${\cal A}$ be a von Neumann algebra of operators on some
finite dimensional Hilbert space.
By a trace on ${\cal  A}$
we shall mean a linear functional $\ta$ which is positive in the sense that
$\ta(A^*A) > 0$
for all non zero $A$ in ${\cal A}$, and cyclic in the sense that
$\ta(AB) = \ta(BA)$
for all $A$ and $B$ in ${\cal A}$. Such a functional is evidently unitarily
invariant in the sense that whenever $A$ and $U$ belong to ${\cal A}$,
and $U$ is unitary, then
$\ta(U^*AU) = \ta(A)$.
Since $\ck$ is a full matrix algebra, it contains all unitaries. Hence any
trace on $\ck$ must assign the same value to all rank one projections, and
thus must be a scalar multiple of the standard trace $Tr$ on the matrix
algebra.
Henceforth, $\ta$ shall denote this trace normalized by the condition that
$\ta(I) = 1$
and $Tr$ shall denote the standard unnormalized trace.

In the non-commutative integration theories of Dixmier [Di53] and Segal [Se53],
the trace functional $\ta$ is the non-commutative analog of the functional
that assigns to an integrable function its integral. When the Hilbert space
is infinite dimensional, some further regularity properties are required of
$\ta$ in order to obtain a useful analog. Since all of our estimations will
be carried out in the finite dimensional setting, we shall not go into this
here, but shall simply refer the reader to these original papers as well as
the accounts in [Gr72] and [Ne74].

Norms on $\ck$ which are the non commutative analogs of
the $L^p$ norms can now be introduced; namely for $1\le p <\infty$ we put
$$\|A\|_p = \bigl(\ta\bigl((A^*A)^{p/2}\bigr)\bigr)^{1/p}\quad,\eqno(3.1)$$
and denote the operator norm of $A$ by $\|A\|_\infty$.

$\ckp$ shall denote $\ck$ equipped with the norm $\|\cdot\|_p$; evidently
$\ckt$ is the Hilbert space of $2^n\times 2^n$ matrices equipped with the
Hilbert-Schmidt norm.

Consider the monomials
$$E_{[\alpha_1\dots,\alpha_j;\beta_1,\dots,\beta_k]} =
Q_{\alpha_1}\cdots Q_{\alpha_j}P_{\beta_1}\cdots P_{\beta_k}\eqno(3.2)$$
where
$\alpha_1>\cdots>\alpha_j$ and $\beta_1>\cdots
>\beta_k$ and $j+k>0$.
Evidently
$$E_{[\alpha_1\dots,\alpha_j;\beta_1,\dots,\beta_k]}^*
E_{[\alpha_1\dots,\alpha_j;\beta_1,\dots,\beta_k]}^{\phantom{.}} = I$$ and thus
$\|E_{[\alpha_1\dots,\alpha_j;\beta_1,\dots,\beta_k]}\|_p = 1$
for all $p$. Moreover
$$\ta(
E_{[\alpha_1\dots,\alpha_j;\beta_1,\dots,\beta_k]}) = 0\quad.\eqno(3.3)$$
To see this, first consider the case in which $j+k$ is odd. The inversion
$x\mapsto -x$ on $\K$ is orthogonal. Hence it induces an automorphism
of $\ck$, and hence there is an invertible $S$ in $\ck$ so that
$$-E_{[\alpha_1\dots,\alpha_j;\beta_1,\dots,\beta_k]} =
SE_{[\alpha_1\dots,\alpha_j;\beta_1,\dots,\beta_k]}S^{-1}\quad.$$
Then, by using cyclicity of the trace we get the desired result. Next
consider the case in which $j+k$ is even and, say, $j>0$. Then write
$E_{[\alpha_1\dots,\alpha_j;\beta_1,\dots,\beta_k]} = Q_{\alpha_1}X$,
and note that by (2.1),
$Q_{\alpha_1}X =
-XQ_{\alpha_1}$. Again the desired conclusion follows from cyclicity of the
trace.

It is easy to see from this that
$$\ta(
E^*_{[\alpha_1\dots,\alpha_j;\beta_1,\dots,\beta_k]}
E^{\phantom .}_{[\gamma_1\dots,\gamma_m;\delta_1,\dots,\delta_n]}) = 0\eqno
(3.4)$$
unless the two monomials coincide. Thus,
together with the identity, the assemblage of such
monomials forms an orthonormal
basis for $\ckt$. Finally observe that since $Qe_- = e_+$
$$\langle \O,
E_{[\alpha_1\dots,\alpha_j]}\O\rangle =0\eqno(3.5)$$
whenever $j \ge 1$, and, as indicated, $k=0$. It now follows that,
restricted to $\cq$,
$$\ta(A) = \langle \O,A\O\rangle\eqno(3.6)$$
for all $A$ in $\ck$.

Formula (3.6) is very important for us. It permits us to calculate
the ``physically'' relevant quantity
$\langle \O,A\O\rangle$ in terms of the apparently mathematically simpler
quantity $\tau(A)$.

Many familiar inequalities for $L^p$ norms hold for the ${\cal C}^p$
norms as well [Di53]. This is true in particular of the H\"older inequality
$$\|AB\|_r \le \|A\|_p\|B\|_q\qquad{1\over r} = {1\over p} + {1\over
q}\quad.$$

Certain optimal inequalities expressing the uniform convexity properties of
the $L^p$ norms also hold for the
${\cal C}^p$ norms, and this fact constitutes one cornerstone of our
analysis.

The modulus of convexity $\delta_p$ of $\ckp$ is defined by
$$\delta_p(\epsilon) =
\inf\bigl\{1 - {1\over 2}\|A+B\|_p : \|A\|_p = \|B\|_p = 1\ ,\ \|A-B\|_p =
\epsilon\bigr\}\eqno(3.7)$$
for $0 < \epsilon < 2$. For $1 < p < \infty$, $\delta_p$ is always positive
which means these norms are uniformly convex. Useful geometric information
is contained in the rate at which $\delta_p(\epsilon)$ tends to zero with
$\epsilon$. It is known [TJ74] that for $2 \le p < \infty$, $\delta_p(\epsilon)
\sim \epsilon^p$, but that for $1 < p \le 2$, $\delta_p(\epsilon) \sim
\epsilon^2$.

An optimal expression of this fact is given by the following theorem which
was proved jointly with Keith Ball [BCL]:
\vskip .3 true in
\noindent{\bf Theorem 1: (Optimal 2-uniform convexity for matrices)}.\quad
{\it For all $m\times m$ matrices $A$ and $B$ and all $p$ for $1 \le p \le
2$,
$$\biggl({Tr|A+B|^p + Tr|A-B|^p\over 2}\biggr)^{2/p} \ge
\bigl(Tr|A|^p\bigr)^{2/p} +
(p-1)\bigl(Tr|B|^p\bigr)^{2/p}\quad.\eqno(3.8)$$
For $ 1 < p < 2$, there is equality only when $B=0$}.
\vskip .3 true in

This result, which we interpret here as a statement about $\ckp$,
is proved in the appendix in the special case that both $A+B$
and $A-B$ are positive; this is the only case in which we shall use it
here, and the proof is considerably simpler
in this case. The full result is proved in
[BCL], in which other geometric inequalities for trace norms are proved as
well.

The theorem implies that
$$\delta_p(\epsilon) \ge
{(p-1)\over 2}\bigl({\epsilon\over 2}\bigr)^2
\qquad{\rm for}\qquad 1 < p \le 2 \eqno(3.9)$$
as one sees by considering
$A = (C+D)/2$, $B = (C-D)/2$, $\|C\|_p = \|D\|_p = 1$ and $\|C - D\|_p =
\epsilon$. It is easily seen that the constant $(p-1)/8$ cannot be
improved.

We make our main application of this result in Section V.
There we will also need to know that the norms on $\ckp$ are continuously
differentiable away from the origin for $1 < p  < \infty$. This is known
[Gr75],
but a simple proof can be based on inequalities of the form
$\delta_p(\epsilon) \ge K_p\epsilon^{r(p)}$ such as we have found above for
$1 < p < 2$. This proof, moreover, gives the modulus of continuity of the
derivative, and is sketched in the appendix as well. Again, these estimates
are independent of the dimension and therefore apply to the case of
infinitely many degrees of freedom.
\vskip .3 true in
\centerline{\bf IV. CONDITIONAL EXPECTATIONS AND THE FERMION}
\centerline{\bf OSCILLATOR SEMIGROUP}
\vskip .3 true in
We are particularly concerned with the subalgebra $\cq$ of $\ck$, and the
conditional expectation [Di53][Um54]
with respect to it shall play a basic role in our
investigation. For any $A$ in $\ck$,
the {\bf conditional expectation} $\pq(A)$
of $A$ with respect to $\cq$ is defined to be the unique
element of $\cq$ such that
$\ta(B^*\pq(A)) = \ta(B^*A)$
for all $B$ in $\cq$. Otherwise said, $\pq$ is the orthogonal projection from
$\ckt$ onto $\cqt$. It is well known that the conditional expectation is
positivity preserving; a familiar argument shows that
$\pq(A^*A) \ge \pq(A)^*\pq(A)$.

We can use the conditional expectation to give a useful expression for the
oscillator semigroup for fermion fields.

Let $\Rt$ be the orthogonal
transformation of ${\cal K}$ given by
$$\Rt(q_j) = (\cos\theta)q_j + (\sin\theta)p_j\eqno(4.1)$$
for each $j$. Of course $\Rt$ gives the evolution at time $\theta$ on phase
space ${\cal K}$ generated by the classical oscillator Hamiltonian
$H({\bf p},{\bf q}) = \sum_{j=1}^n{1\over 2}(p_j^2 + q_j^2)$.
Let $\Rt$ denote
the automophism of $\ck$ generated by the orthogonal transformation $\Rt$
as in the first section.
For each $t\ge 0$, define $\theta(t) = \arccos(e^{-t})$ and define the
operator $P_t$ on $\ckt$ by
$$P_tA = \pq\circ\rt\circ\iq A\quad.\eqno(4.2)$$
Note that $\iq$ is the {\bf natural imbedding} of $\cq$ into $\ck$,
and regarded as such, it is a $*$-automorphism.
Formula (4.2) is the analog of the familiar expression for the
boson oscillator semigroup on $L^2(\Q,(2\pi)^{-n/2}e^{-q^2/2}{\rm d}^nq)$, i.e.
the Mehler semigroup
$$P_t^{(boson)}A({\bf q}) = \int_\Q A\biggl(e^{-t}{\bf q} + (1-e^{-2t})^{1/2}
{\bf p}\biggr)
(2\pi)^{-n/2}e^{-p^2/2}{\rm d}^np\quad.$$

Note that since all of the operators on the right in (4.2) are positivity
preserving, so is $P_t$. Also, since the first two operations on the right
preserve the ${\cal C}^p$-norms, and since the conditional
expectation is readily seen to
be a contraction from $\ckp$ to $\cqp$ for each $p$, it is readily seen that
$P_t$ possesses this property as well.

To obtain a more familiar expression for $P_t$, note that
$$R_{\theta(t)}\circ\iq\bigl(E_{[\alpha_1,\dots,\alpha_k]}\bigr)
= e^{-kt}E_{[\alpha_1,\dots,\alpha_k]} + \bigl({\rm terms\ annihilated\ by}\
\pq\bigr)\quad.$$ Hence
$$P_t\bigl(E_{[\alpha_1,\dots,\alpha_k]}\bigr)
= e^{-kt}E_{[\alpha_1,\dots,\alpha_k]}.\eqno(4.3)$$
Evidently $\{P_t : t\ge0\}$ is generated by $H_0$ where
$H_0\bigl(E_{[\alpha_1,\dots,\alpha_k]}\bigr) =
kE_{[\alpha_1,\dots,\alpha_k]}$.
It is easy to see that under the unitary equivalence between $\cqt$
and fermion Fock space ${\cal F}$ described by Segal [Se56], $H_0$
is equivalent to the usual number operator, or in other words, oscillator
Hamiltonian on ${\cal F}$.

Our primary goal is to prove optimal hypercontractivity bounds for $P_t$.
That is,
given $1 < p < q <\infty$ we want to show that for some finite $t$, $P_t$ is a
contraction from $\cqp$ to $\cqq$, and to find the smallest such $t$. Let
$$\|P_t\|_{p\rightarrow q}^{\phantom .} =
\sup\{\|P_tA\|_q\ :\ \|A\|_p = 1\ \}\quad.
\eqno(4.4)$$
As a first reduction, we shall show that the supremum on the right in (4.4)
can be restricted to the positive operators $A$ with $\|A\|_p = 1$.
In the boson case this follows immediately from the fact that, in
ordinary probability theory, the absolute value of a conditional expectation
is no greater than the conditional expectation of the absolute value.

In general, matters concerning the absolute value in the non-commutative
setting are more troublesome than in the commutative setting. An
example is provided by the Araki-Yamagami inequality [ArYa] which,
specialized to our context, asserts
that the map $A\mapsto |A|$ is Lipschitz continuous on $\ckt$ with constant
$\sqrt 2$ instead of the constant 1 which we would have in the commutative
setting.

Thus while the conditional expectation in an
operator algebra has many properties
analogous to those of the conditional expectation in ordinary probability
theory [Um54], it is not in general true that $|\pq(A)|$ will be a
smaller operator than $\pq(|A|)$. The following theorem expresses a useful
property in this direction which does hold, and after proving it we shall
show by example
that stronger properties do not hold. The theorem and its proof are easily
extended to a more general von Neumann algebra setting by the methods in
[Ru72].
\vskip .3 true in
\noindent{\bf Theorem 2: (A Schwarz inequality for conditional expectations)}.
\quad {\it For all $A$ in $\ck$ and all $p$ with} $1\le
p\le\infty$,
$$\|\pq(A)\|_p \le \|\pq(|A|)\|_p^{1/2}
\|\pq(|A^*|)\|_p^{1/2}\quad.\eqno(4.5)$$
\vskip .3 true in
\noindent{\bf Remark:}\ If we let $F(A)$ denote $\|\pi_{\cal Q}A\|_p$,
then the same argument which we shall use to prove
Theorem 2 also establishes that
$$F(A^*B) \le F(A^*A)^{1/2}F(B^*B)^{1/2}\quad.\eqno(4.6)$$
In this form, the term ``Schwarz inequality'', by which we referred to
(4.5), is more evidently appropriate. Moreover, inequalities of the
type (4.6) are well known in matrix analysis for many familiar functions;
for example when F(A) is the determinant of $A$ or the spectral radius of
$A$. Further examples can be found in [MeDS].

In [Li76] it is shown that  for a function $F$
that satisfies (4.6), and which is monotone
increasing; i.e. satisfies $F(B) \ge F(A)$ for all $B\ge A\ge 0$,
the following inequalities hold:
$$F(\sum_{j=1}^mA_j^*B_j) \le F(\sum_{j=1}^m(A_j^*A_j)^{1/2}
F(\sum_{j=1}^m(B_j^*B_j)^{1/2}$$
and
$$F(\sum_{j=1}^mA_j) \le F(\sum_{j=1}^m(|A_j|)^{1/2}
F(\sum_{j=1}^m(|A^*_j|)^{1/2}\quad.$$
In particular, these inequalities hold for $F(A) =
\|\pi_{\cal Q}A\|_p$.

Specializing the last inequality to the case $m=1$ then yields (4.5).
In our present case however, the proof of the (4.6) is essentially
the same as the direct proof of (4.5). Nonetheless,
it should not be considered novel that by taking $|A^*|$
into consideration as well as $|A|$, we can obtain a suitable bound on
$\|\pi_{\cal Q}A\|_p$.
\vskip .3 true in

\noindent{\bf Proof:}\
Let $A = U|A|$ be the polar decomposition of  $A$.
Then $\|\pq(A)\|_p = \ta(CU|A|)$ for some $C$ in
$\cq$ with $\|C\|_{p'} = 1$. Let $C = V|C|$ be the polar decomposition of
$C$. Both  $V$ and $|C|$ belong to $\cq$ as well. Thus
$$\|\pq(A)\|_p = \ta\bigl(CU|A|^{1/2}|A|^{1/2}\bigr)
= \ta\bigl(|C|^{1/2}U|A|^{1/2}|A|^{1/2}V|C|^{1/2}\biggr)$$
$$\le \ta\bigl(|C|^{1/2}U|A|U^*|C|^{1/2}\bigr)^{1/2}
\ta\bigl(|C|^{1/2}V^*|A|V|C|^{1/2}\bigr)^{1/2}$$
$$=\ta\bigl(|C|(U|A|U^*)\bigr)^{1/2}
\ta\bigl((V|C|V^*)|A|\bigr)^{1/2}
\le\|\pq(U|A|U^*)\|_p^{1/2}
\|\pq(|A|)\|_p^{1/2}\quad.$$
Finally, we note that $U|A|U^* = |A^*|$.
\quad$\square$

\vskip .4 true in
\noindent{\bf Example}\quad Let $A$ be the matrix $A = \left[\matrix{0&1\cr
0&1\cr}\right]$. Then
$$|A| = \sqrt2\left[\matrix{0&0\cr0&1}\right]\qquad{\rm and}\qquad
|A^*| = {1\over \sqrt2}\left[\matrix{1&1\cr1&1}\right]\quad.$$
Note that $|A^*|$ is in $\cq$, but $A$ and $|A|$ are not. One easily finds
$$\pq(A) =
{1\over 2}\left[\matrix{1&1\cr1&1}\right]\quad,\quad \pq(|A|) =
{1\over \sqrt2}\left[\matrix{1&0\cr0&1}\right]\quad,\quad
\pq(|A^*|) =
{1\over \sqrt2}\left[\matrix{1&1\cr1&1}\right]\ .$$
$$\|\pq(A)\|_\infty = 1\qquad,\qquad \|\pq(|A|)\|_\infty = {1\over
\sqrt2}\qquad{\rm and}\qquad
\|\pq(|A^*|)\|_\infty = \sqrt2\quad.$$
\vskip .4 true in
\noindent{\bf Theorem 3: ($P_t$ has positive maximizers)}.\quad
{\it The norm of $P_t$ from $\cqp$ to $\cqq$ is achieved on the positive
operators; i.e.}
$$\|P_t\|_{p\rightarrow q} =
\sup\{\|P_tA\|_q\ :\ A\ge 0\quad{\rm and}\quad\|A\|_p = 1\ \}\quad.\eqno(4.7)$$
\vskip .4 true in
\noindent{\bf Proof:}\quad Since $R_\theta$ and $\iq$ are both
$*$-automorphisms, $|\rt\circ\iq A| = \rt\circ\iq|A|$ and
$|(\rt\circ\iq A)^*| = \rt\circ\iq |A^*|$. Thus
$\|P_tA\|_q \le \|P_t|A|\|_q^{1/2}\|P_t|A^*|\|_q^{1/2}$,
and of course $\||A^*|\|_p = \||A|\|_p = \|A\|_p$.
\quad $\square$
\vskip .4 true in
\centerline{\bf V. HYPERCONTRACTIVITY FOR FERMIONS}
\vskip .3 true in

Our main result is the following theorem which is established in this section.
\vskip .3 true in
\noindent{\bf Theorem 4: (Optimal fermion hypercontractivity)}.\quad {\it For
all $1 < p \le q < \infty$, $\|P_t\|^{\phantom .}_{p\rightarrow
q} = 1$ exactly when $$e^{-2t} \le {p-1\over q-1}\quad.$$}
\vskip .3 true in

The heart of the matter is the following lemma:
\vskip .3 true in
\noindent{\bf Lemma:}\quad {\it For all $1 < p \le 2$, \ $\|P_t\|_{p\rightarrow
2} = 1$ exactly when $e^{-2t} \le (p-1)$.}
\vskip .3 true in
\noindent{\bf Proof:}\quad Fix a positive element A of $\cq$.
Pick a basis $\{q_1,\dots,q_n\}$ of $\Q$, and let
$\cqn$ denote the Clifford algebra associated with the span of the first
$n-1$ of these basis elements. It is evident from the form of our standard
basis of $\cq$ that $A$ can be uniquely decomposed as
$A = B + CQ_n$
where $B$ and $C$ belong to $\cqn$. Then using the Jordan-Wigner transform
we can write
$A = B + CV_n\hat Q_n$.
Now write
$$\H = \H_{(n-1)}\otimes\IC^2\eqno(5.1)$$
so that $B$, $C$ and $V_nC$ can be
considered as operators on the first factor ${\cal H}_{(n-1)}$.
Let $u_\pm = (e_+ \pm
e_-)/\sqrt 2$, so that $Qu_\pm = \pm u_\pm$. Then if $v$ is any
vector in $\H_{(n-1)}$,
$$\langle (v\otimes u_\pm),A(v\otimes u_\pm)\rangle_{\H}^{\phantom .} =
\langle v,\bigl(B \pm CV_n\bigr)v\rangle_{\H_{(n-1)}}^{\phantom .}
\quad.\eqno(5.2)$$
We see from this that since $A\ge0$, so are both $B + CV_n\ge 0$ and
$B - CV_n\ge 0$. Now let $Tr_1$  and $Tr_2$ denote the partial traces
over the first and second factors in (5.1), so that with $Tr$ still denoting
the full trace,
we have $Tr = Tr_1Tr_2$. Now applying Theorem 1 in the special case
which is proved in the appendix
$$\|A\|_p^2 = \biggl({1\over 2^n}Tr|B + CV_n\hat Q_n|^p\biggr)^{2/p}$$
$$= \biggl({1\over 2^{(n-1)}}\biggr)^{2/p}
\biggl({Tr_1|B + CV_n|^p + Tr_1|B - CV_n|^p\over 2}\biggr)^{2/p}$$
$$\ge \biggl({1\over 2^{(n-1)}}\biggr)^{2/p}
\biggl((Tr_1|B|^p)^{2/p} + (p-1)(Tr_1|C|^p)^{2/p}\biggr)$$
since $V_n$ is unitary.

Thus
$\|A\|_p^2 \ge \|B\|_p^2 + (p-1)\|C\|_p^2$
where the norms on the right are all norms on $\cqn$.

Now we make the inductive assumption that the lemma has been established
for $\cqn$. This is clearly the case when $n=1$. We then have
$$\|A\|_p^2 \ge \|P_tB\|_2^2 + (p-1)\|P_tC\|_2^2$$
where $e^{-2t} = (p-1)$. But clearly from (4.3)
$\|P_tCQ_n\|_2^2 = (p-1)\|P_tC\|_2^2$
where the norms are once again norms on $\cq$. Moreover by (3.4) $P_tB$ and
$P_tCQ_n$ are orthogonal. Thus
$\|P_tB\|_2^2 + (p-1)\|P_tC\|_2^2 =
\|P_t(B + CQ_n)\|_2^2 = \|P_tA\|_2^2$.\quad
$\square$
\vskip .3 true in
\noindent{\bf Theorem 5: (Optimal fermion logarithmic Sobolev inequality)}.\
{\it For all} $A \in \cq$,
$$\ta\bigl(|A|^2\ln|A|^2\bigr) -
\bigl(\|A\|_2^2\ln\|A\|_2^2\bigr) \le 2\langle A,H_0A\rangle\quad.\eqno(5.3)$$
\vskip .3 true in
\noindent{\bf Proof:}\ By the lemma, $\|P_tA\|_2^2 \le \|A\|^2_{(1+e^{-2t})}$
and there is equality at $t=0$. Both sides are continuously differentiable,
and comparing derivatives at $t=0$ we obtain the result. Indeed,
$${{\rm d}\over {\rm d}p}\|A\|_p = {1\over p}\|A\|_p^{1-p}\biggl(
\ta(|A|^p\ln|A|) - \|A\|_p^p\ln\|A\|_p\biggr) \eqno(5.4)$$
and of course
$${{\rm d}\over {\rm d}t}\|P_tA\|_2^2 =
{{\rm d}\over {\rm d}t}\langle A,P_{2t}A\rangle =
-2\langle A,H_0A\rangle\quad.\ \square$$
\vskip .5 true in

Gross refers to the quadratic form on the right side of (5.3) as the
Clifford Dirichlet form since it shares many properties of Dirichlet forms
in the ordinary commutative setting. An approach to the development of a
theory of Dirichlet forms in the non-commutative setting can be found in
[AlHK].
\vskip .5 true in
\noindent{\bf Proof of Theorem 4:}\quad By a deep result of Gross, when $A\ge0$
and $1 < p < \infty$,
$$\langle A^{p/2},H_0A^{p/2}\rangle \le
{(p/2)^2\over p-1}\langle A,H_0A^{p-1}\rangle \quad.$$
Replacing $A$ in (5.3) by $A^{p/2}$ and using the inequality just quoted
we obtain, following Gross's ideas [Gr75],
$$\ta(A^p\ln A) - \|A\|_p^p\ln\|A\|_p \le {p/2\over p-1}\langle
A^{p-1},H_0A\rangle\quad.$$
By combining this with (5.4) a differential inequality is obtained which
implies
that
$\|P_tA\|_{q(t)}$
is a decreasing function of $t$
when $q(t) = 1 + e^{2t}(p-1)$.
This establishes the result for $A\ge0$,
and by Theorem 3 it is established in general.
By (4.3), $P_tI = I$, and therefore $\|P_t\|_{p\rightarrow q}$ is always
at least 1 for all $p$ and $q$.
That the inequality is best
possible follows from a direct computation with one degree of freedom.
To be precise, $\|P_t(I+Q_1)\|_q = \|I+e^{-t}Q_1\|_q$ is easily computed
and compared with $\|I+Q_1\|_p$ [Gr72]. The first quantity is greater than
the second if $e^{-2t} > (p-1)/(q-1)$.
\quad $\square$
\vskip .3 true in

\noindent{\bf VI. HYPERCONTRACTIVITY FOR BOSONS AND FERMIONS TOGETHER}
\vskip .3 true in

As a result of the present work and of earlier work on bosons, we know
that for $t$ given by $e^{-2t} = (p-1)/(q-1)$, both the fermion and the
boson oscillator semigroups are contractive from the appropriate
$p$--spaces to the appropriate $q$--spaces, and that this value of
$t$ is optimal for each case {\it separately}.

It is natural to expect that the same condition governs hypercontractivity
in a situation in which we have bosons and fermions {\it together}. This
is indeed the case, as we now show using Minkowski's inequality in an
argument based on Segal's method for showing that the optimal conditions
for hypercontractivity with $m$ boson degrees of freedom are the same as
for one degree of freedom.

Let $\mu({\rm d}x) = (2\pi)^{-m/2}e^{-x^2/2}{\rm d}x$ be the unit Gauss measure
on $\IR^m$. Then in our mixed setting, with $m$ boson degrees of freedom
and $n$ fermion degrees of freedom, the relevant $p$--space is
$${\cal B}^p = L^p(\IR^m,\mu)\otimes{\cal C}^p({\cal Q}_{(n)}),$$
which may be regarded as consisting of
${\cal C}^p({\cal Q}_{(n)})$ valued measurable functions $x\mapsto A(x)$
such that
$$|\!|\!|A|\!|\!|_p^p = \int_{\IR^m}\|A(x)\|_p^p\mu({\rm d}x)$$
is finite. This equation defines the norm on ${\cal B}^p$.
For $p=2$, ${\cal B}^p$ is naturally isomorphic to the tensor product of
the symmetric tensor algebra over $\IC^m$ and the antisymmetric tensor
algebra over $\IC^n$ as shown by Segal. On the latter space we have the
mixed oscillator semigroup generated by the sum of the boson and fermion
number operators
$$\exp\biggl\{-t\biggl[\sum_{j=1}^m a^*_ja_j+\sum_{j=1}^n c^*_jc_j
\biggr]\biggr\}.$$
Considered as operators on ${\cal B}^p$, the operators
${\cal P}_t$ which constitute this semigroup are given by
$${\cal P}_tA(x) =
\int_{R^M}M_t(x,x')P_t\bigl[A(x')\bigr]\mu({\rm d}x')\eqno(6.1)$$ where
$M_t(x,x')$ is the Mehler kernel; i.e., the positive integral kernel for the
boson oscillator semigroup $P^{(boson)}_t$ discussed in Section IV.
Of course $P_t$ denotes the fermion oscillator semigroup studied throughout
this paper.

Now successively applying Minkowski's inequality, our theorem on optimal
fermion hypercontractivity, and Nelson's theorem on optimal boson
hypercontractivity, we have for $e^{-2t} \le (p-1)/(q-1)$:

$$\eqalign{|\!|\!|{\cal P}_tA|\!|\!|_q^q&=
\int_{\IR^m}\bigg\|\int_{\IR^m}M_t(x,x')P_t\bigl[A(x')\bigr]
\mu({\rm d}x')
\bigg\|_q^q\mu({\rm d}x)\cr
&\le \int_{\IR^m}\biggl(\int_{\IR^m}M_t(x,x')\|P_tA(x')\|_q\mu({\rm
d}x')\biggr)
^q\mu({\rm d}x)\cr
&\le \int_{\IR^m}\biggl(\int_{\IR^m}M_t(x,x')\|A(x')\|_p\mu({\rm
d}x')\biggr)^q\mu({\rm d}x)\cr
&\le \bigl(\int_{\IR^m}\|A(x)\|_p^p\mu({\rm d}x)\bigr)^{q/p} =
|\!|\!|A|\!|\!|_p^q \quad.\cr}$$

\vskip .3 true in
\centerline{\bf APPENDIX}
\vskip .3 true in

\noindent{\bf Proof of Theorem 1 when $A\pm B\ge0$:}\quad
Let $Z$ and $W$ be the $2m\times 2m$ matrices given by
$$Z = \left[\matrix{A&0\cr 0&A\cr}\right]\quad,\quad W =
\left[\matrix{B&0\cr 0&-B\cr}\right]\quad.$$
Our goal is to establish that for all $r$ with $0\le r \le 1$,
$$\biggl({Tr(A+rB)^p + Tr(A-rB)^p\over 2}\biggr)^{2/p} \ge (Tr(A)^p)^{2/p} +
r^2(p-1)(Tr|B|^p)^{2/p}\quad,$$
or what is the same,
$$Tr(Z+rW)^{2/p} \ge (Tr(Z)^p)^{2/p} + r^2(p-1)(Tr|W|^p)^{2/p}
\quad.\eqno(A.1)$$
First, note that the null space of $Z+rW$ is exactly the null space of $Z$ for
$0 \le r < 1$. Thus by carrying out all of the following computations on
the orthogonal complement of this fixed null space, we may freely assume
that $Z+rW > 0$ for all $0 \le r < 1$.
Next, both sides of (A.1) agree at $r=0$,
and the first derivatives in $r$ of both
sides vanish there as well. We define $\psi(r)$ to be $Tr(Z+rW)^p$. Then the
second derivative in $r$ of the left side of (A.1) satisfies
$${{\rm d}^2\over {\rm d}r^2}\bigl(\psi(r)\bigr)^{2/p} \ge
{2\over p}\psi(r)^{(2-p)/p}{{\rm d}^2\over {\rm d}r^2}\psi(r)\quad.$$
The second derivative on the right side is just
$$2(p-1)(Tr|W|^p)^{2/p}\quad,$$
and we are left with showing that
$${1\over p}\psi(r)^{(2-p)/p}
{{\rm d}^2\over {\rm d}r^2}\psi(r) \ge (p-1)(Tr|W|^p)^{2/p} \eqno(A.2)$$
for all $0 < r < 1$. By redefining $Z$ to be $Z+rW$, it suffices to
establish (A.2) at $r=0$.

Now
${{\rm d}\over {\rm d}r}\psi(r) =
p\bigl(Tr(Z+rW)^{(p-1)}W\bigr)$,
since $A\pm B\ge 0$, $Z+rW\ge0$ for small $r$, and
we can use the integral representation
$$\bigl(Z+rW\bigr)^{(p-1)} = c_p\int_0^\infty t^{(p-1)}\biggl[{1\over t} -
{1\over t + (Z+rW)}\biggr]dt$$ to conclude that
$${{\rm d}^2\over {\rm d}r^2}\psi(0) =
pc_p\int_0^\infty t^{(p-1)}Tr\biggl[
{1\over t + Z}W
{1\over t + Z}W\biggr]dt\quad. \eqno(A.3)$$

Consider the right side as a function, $f(Z)$, of $Z$ for fixed $W$. It is easy
to
see that $f$ is convex in $Z$. (Simply replace $Z$ by $Z+tX$, with $X$
self--adjoint, and then differentiate twice with respect to $t$; the
positivity follows from the Schwarz inequality for traces.) Also,
$f(UZU^*)=f(Z)$
provided $U$ is unitary and $U$ commutes with $W$.
In a basis in which $W$ is diagonal, we form the set $\cal{U}$ consisting
of the $2^{2m}$ distinct
diagonal unitary
matrices, each with $+1$ or $-1$ in each diagonal entry. Each of these
clearly commutes with $W$. Then
$$f(Z)=2^{-2m}\sum_{U\in\cal{U}}f(UZU^*)\ge f(2^{-2m}\sum_{U\in\cal{U}}UZU^*)
=f(Z_{{\rm diag}}),$$
where
$Z_{{\rm diag}}$ is the matrix that is diagonal in the basis diagonalizing
$W$, and whose diagonal entries are those of $Z$ in this
basis.

Replacing $Z$ by
$Z_{{\rm diag}}$ in (A.3), the integration can be carried out, and we
obtain
$${{\rm d}^2\over {\rm d}r^2}\psi(0) \ge p(p-1)
\biggl(\sum_{j=1}^{2m}z_j^{(p-2)}w_j^2\biggr)$$
where $z_j$ and $w_j$, respectively, denote
the $j$th diagonal entries of $Z$ and $W$ in a
basis diagonalizing $W$.

Now consider $\psi(0) = Tr(Z^p)$ as a function of $Z$. It is clearly
convex,
and thus by the averaging method just employed, we obtain
$$\psi(0) \ge
\biggl(\sum_{j=1}^{2m}z_j^p\biggr)\quad.$$
To establish (A.2), we are only left with showing that
$$\biggl(\sum_{j=1}^{2m}z_j^p\biggr)^{(2-p)/p}
\biggl(\sum_{j=1}^{2m}z_j^{(p-2)}w_j^2\biggr) \ge
\biggl(\sum_{j=1}^{2m}|w_j|^p\biggr)^{2/p}\quad,\eqno(A.4)$$
but this follows immediately from H\"older's inequality.

To complete the proof, observe that equality in (A.1) for $r=1$ and $1 < p
< 2$ implies equality in
(A.4) for almost every $r$ in $[0,1]$. Here, recall that $z_j$ in (A.4)
really denotes the $j$th diagonal element of $Z+rW$; these are the numbers
$z_j+rw_j$, where $z_j$ denotes the $j$th diagonal element of $Z$.

Let us assume that $w_j \neq 0$ for some $j$. Then
equality in H\"older's inequality (A.4)
requires that the vector with positive components
$z_j + rw_j$ be proportional to the vector with components $|w_j|$. Thus,
for almost every $r$ in $[0,1]$ we require
$$z_j + rw_j = c(r)|w_j|$$
for some number $c(r)$ that depends on $r$ but not on $j$. The left side
above is a linear function, and thus $c(r) = a+rb$ for some numbers $a$ and
$b$.
But then clearly $b = w_j/|w_j|$, and all
non-zero
eigenvalues of $W$ would necessarily
have the same sign. This is impossible since $TrW
= 0$.
\quad $\square$
\vskip .3 true in

We now give an application of the uniform convexity implied by this theorem
to the differentiability of the $\ckp$ norms. First we recall that for
$2 \le p < \infty$, the modulus of convexity is given by an analog of an
inequality of Clarkson for integrals which Dixmier [Di53]
established for traces. Specializing to $\ckp$, this inequality reads
$$\biggl\|{A+B\over 2}\biggr\|_p^p +
\biggl\|{A-B\over 2}\biggr\|_p^p
\le {1\over 2}\bigl[\|A\|^p_p + \|B\|_p^p\bigr]\qquad
2\le p < \infty\quad,$$
which implies that in this range ({\it cf}. (3.7))
$$\delta_p(\epsilon) \ge {1\over p}\bigl({\epsilon\over 2}\bigr)^p
\quad.\eqno(A.5)$$

For any non-zero $A$ in $\ckp$, define $\da$ by
$$\da = \|A\|_p^{(1-p)}|A|^{(p-1)}U^*$$
where $A = U|A|$ is the polar decomposition of $A$. Let $p'$ be defined by
$1/p + 1/p' = 1$.  Then for $1<p<\infty$, $\|\da\|_{p'} = 1$
and $\ta(\da A) = \|A\|_p$. Moreover, $\da$ is the unique element of
${\cal C}^{p'}({\cal K})$ with this property. We call the map $A\mapsto
\da$ the {\bf gradient map} on $\ckp$. The next theorem sharpens a result
of Gross [Gr75].
\vskip .3 true in
\noindent{\bf Theorem 6: (H\"older continuity of the gradient map on $\ckp$)}.
\quad
{\it For all $1 < p <\infty$, the gradient map is norm continuous.
Moreover,
$$\|\da - \db\|_{p'} \le 2\biggl(p'{\|A-B\|_p\over \|A+B\|_p}\biggr)^{(p-1)}
\quad for\quad 1<p\le 2\quad,
\eqno(A.6)$$
and
$$\|\da - \db\|_{p'} \le 4(p-1)\biggl({\|A-B\|_p\over \|A+B\|_p}\biggr)
\quad for\quad 2\le p<\infty\quad.
\eqno(A.7)$$}
\vskip .3 true in
\noindent{\bf Proof:}\quad First observe that for all $1 \le p \le \infty$
$$\|\da + \db\|_{p'}\|A+B\|_p \ge \Re\ta\bigl((\da+\db)(A+B)\bigr)$$
$$= 2\bigl(\|A\|_p + \|B\|_p\bigr) - \Re\ta\bigl((\da - \db)(A-B)\bigr)$$
$$\ge 2\bigl(\|A + B\|_p\bigr) - \|\da - \db\|_{p'}\|A-B\|_p\quad.$$
Thus,
$$1 - \biggl\|{\da + \db\over 2}\biggr\|_{p'}
\le \biggl\|{\da - \db\over 2}\biggr\|_{p'}\biggl({\|A-B\|_p\over \|A+B\|_p}
\biggr)\quad.\eqno(A.8)$$
But, by (A.5) and (3.7), we have
$$1 - \biggl\|{\da + \db\over 2}\biggr\|_{p'} \ge {1\over p'}
\biggl\|{\da - \db\over 2}\biggr\|_{p'}^{p'}\eqno(A.9)$$
for $1<p\le2$. By combining (A.9) and (A.8) we obtain (A.6). Similarly,
by combining (3.9) with (A.8) we obtain (A.7). \ $\square$
\vskip .3 true in

The continuity of the gradient map for $\ckp$
has been established by Gross [Gr72], but
his proof is more involved and
does not yield an estimate of the modulus of continuity. It is now
easy to establish continuous differentiability of the $\ckp$ norms away
from the origin since, with $h(t) = \|A+tB\|_p$, with $A$ different from $0$
and with $t$ and $s$ sufficiently small, we have
$$\Re\ta\bigl({\cal D}(A+tB)B\bigr) \le {h(t+s) - h(t)\over s} \le
\Re\ta\bigl({\cal D}(A+(t+s)B)B\bigr)\eqno(A.10)$$
when $s$ is positive.
To see this, observe that $h(t) = \Re\tau\bigl({\cal D}(A+tB)(A+tB)\bigr)$,
and that $h(t+s) \ge
\Re\tau\bigl({\cal D}(A+tB)(A+(t+s)B)\bigr)$
by H\"older's inequality. By subtracting the expression
for $h(t)$ from the estimate for $h(t+s)$ and dividing by $s$, we obtain
the inequality on the left in (A.10). The inequality on the right is
obtained in an analogous manner.
When $s$ is negative, the inequalities are clearly reversed.
Letting $s$ tend to zero, we obtain
$${{\rm d}\over {\rm d}t}\|A+tB\|_p\biggl|_{t=0} =
\Re\ta\bigl(\da B\bigr)\quad.$$
\vskip .3 true in
\centerline{\bf REFERENCES}
\vskip .3 true in

\item{[AlHK]} Albeverio, S., H\o egh-Krohn, Dirichlet forms and Markov
semigroups on $C^*$-algebras, Commun. Math. Phys., {\bf 56} (1977) 173-187.

\item{[ArYa]} Araki, H.: Yamagami, S.: An inequality for the Hilbert-Schmidt
norm, Commun. Math. Phys., {\bf 81} (1981) 89-96.

\item{[BCL]} Ball, K., Carlen, E.A., Lieb, E.H.: preprint 1992.

\item{[BrWe]} Brauer, R., Weyl, H.: Spinors in $n$ dimensions, Am. Jour.
Math., {\bf 57} (1935) 425-449.

\item{[CL91]} Carlen, E.A., Loss, M.: Extremals of functionals with competing
symmetries, Jour. Func. Analysis {\bf 88} (1991) 437-456

\item{[Da76]} Davies, E.B.: {\it Quantum  Theory of Open Systems},
Academic Press, New York, 1976.

\item{[Di53]} Dixmier, J.: Formes lin\'eaires sur un anneau d'op\'erateurs,
Bull. Soc. Math. France {\bf 81} (1953) 222-245.

\item{[Fe69]} Federbush, P.: A partially alternate derivation of a result
of Nelson, Jour. Math. Phys., {\bf 10} (1969) 50-52.

\item{[Gr72]} Gross, L.: Existence and uniqueness of physical ground
states, Jour. Funct. Analysis, {\bf 10} (1972) 52-109.

\item{[Gr75]} Gross, L.: Hypercontractivity and logarithmic Sobolev
inequalities for the Clifford-Dirichlet form, Duke Math. J., {\bf 43}
(1975) 383-396.

\item{[Gr89]} Gross, L.: Logarithmic Sobolev inequalities for the heat
kernel on a Lie group and a bibliography on logarithmic Sobolev
inequalities and hypercontractivity, pp. 108-130 in {\it White Noise
Analysis, Mathematics and Applications}, eds. Hida et al., World
Scientific, Singapore 1990.

\item{[HuPa]} Hudson, R, Parthasarathy, K.R.: Quantum It\^ o's formula
and stochastic evolutions, Commun. Math. Phys., {\bf 93} (1984) 301-323.

\item{[JoKl]} Jordan, P., Klein, O.: Z\"um Mehrk\"orperproblem der
Quantentheorie,
Zeits. f\"ur Phys., {\bf 45} (1927) 751-765.

\item{[JoWi]} Jordan, P., Wigner, E.P.: \"Uber das Paulische
\"Aquivalenzverbot,
Zeits. f\"ur Phys., {\bf 47} (1928) 631-651.

\item{[Li76]} Lieb, E.H.: Inequalities for some operator and matrix
functions, Adv. Math. {\bf 20} (1976) 174-178

\item{[Li90]} Lieb, E.H.: Gaussian kernels have only Gaussian maximizers,
Invent. Math. {\bf 102} (1990) 179-208

\item{[Lin]} Lindsay, M.: Gaussian hypercontractivity revisited, Jour.
Funct. Analysis {\bf 92} (1990) 313-324.

\item{[LiMe]} Lindsay, M., Meyer, P.A.: preprint, 1991.

\item{[MeDS]} Merris, R.,Dias da Silva,J.A.: Generalized Schur functions,
Jour. Lin. Algebra {\bf 35} (1975) 442-448.

\item{[Me85]} Meyer, P.A.: El\'ements de probabilit\'es quantiques,
expos\'es I-V, pp.186-312 in
Sem. de Prob. XX, Lecture notes in Math. {\bf 1204},
Springer, New York, 1985.

\item{[Me86]} Meyer, P.A.: El\'ements de probabilit\'es quantiques,
expos\'es VI-VIII, pp. 27-80 in
Sem. de Prob. XXI, Lecture notes in Math. {\bf 1247},
Springer, New York, 1986.

\item{[Ne66]} Nelson, E.: A quartic interaction in two dimensions, in
{\it Mathematical Theory of Elementary Particles}, R. Goodman ans I. Segal
eds., MIT Press, Cambridge Mass., 1966.

\item{[Ne73]} Nelson, E.: The free Markov field, Jour. Funct. Analysis,
{\bf 12} (1973) 211-227.

\item{[Ne74]} Nelson, E.: Notes on non-commutative integration, Jour. Funct.
Analysis, {\bf 15} (1974) 103-116.

\item{[Nev]} Neveu, J.: Sur l'esperance conditionelle par rapport \`a un
mouvement Brownien, Ann. Inst. H. Poincar\'e Sect. B. (N.S.) {\bf 12}
(1976) 105-109

\item{[Ru72]} Ruskai, M.B.: Inequalities for traces on Von Neumann
algebras,
Commun. Math. Phys., {\bf 26} (1972) 280-289.

\item{[SML]} Schultz, T.D., Mattis, D.C., Lieb, E.H.: Two dimensional Ising
model as a soluble problem of many fermions, Rev. Mod. Phys. {\bf 36}
(1964) 856-871.

\item{[Se53]} Segal, I.E.: A non-commutative extension of abstract
integration, Annals of Math., {\bf 57} (1953) 401-457.

\item{[Se56]} Segal, I.E.: Tensor algebras over Hilbert spaces II,
Annals of Math., {\bf 63} (1956) 160-175.

\item{[Se70]} Segal, I.E.: Construction of non-linear local quantum
processes: I, Annal of Math., {\bf 92} (1970) 462-481.

\item{[TJ74]} Tomczak-Jaegermann, N.: The moduli of smoothness and
convexity and Rademacher averages of trace classes $S_p (1\le p
< \infty)$, Studia Mathematica {\bf 50} (1974) 163-182.

\item{[Um54]} Umegaki, H.: Conditional expectation in operator algebras I,
Tohoku Math. J., {\bf 6} (1954) 177-181.

\end